\documentclass[12pt]{article}
\usepackage{amsmath}
\usepackage{latexsym}

\begin{document}

\begin{center}
{\Large \bf  Partner symmetries of the complex Monge-Amp\`ere
equation yield hyper-K\"ahler metrics without continuous symmetries}\\[2mm]
{\large \bf A A Malykh$^1$, Y Nutku$^2$ and
M B Sheftel$^{1,2}$}  \\[1mm]
$^1$ Department of Higher Mathematics, North Western State
Technical University, Millionnaya St. 5, 191186, St. Petersburg,
Russia
\\ $^2$ Feza G\"{u}rsey Institute, PO Box 6, Cengelkoy,
81220 Istanbul, Turkey
\\ E-mail: specarm@online.ru, nutku@gursey.gov.tr,
sheftel@gursey.gov.tr
\end{center}

\vspace{1mm}

\noindent  {\bf Abstract} \\

We extend the Mason-Newman Lax pair for the elliptic complex
Monge-Amp\`ere equation so that this equation itself emerges as an
algebraic consequence. We regard the function in the extended Lax
equations as a complex potential. They imply the determining
equation for symmetries of the complex Monge-Amp\`ere equation as
their differential compatibility condition. We shall identify the
real and imaginary parts of the potential, which we call partner
symmetries, with the translational and dilatational symmetry
characteristics respectively. Then we choose the dilatational
symmetry characteristic as the new unknown replacing the K\"ahler
potential. This directly leads to a Legendre transformation.
Studying the integrability conditions of the Legendre-transformed
system we arrive at a set of linear equations satisfied by a
single real potential. This enables us to construct non-invariant
solutions of the Legendre transform of the complex Monge-Amp\`ere
equation.

Using these solutions we obtained explicit Legendre-transformed
hyper-K\"ahler metrics with anti-self-dual Riemann curvature
$2$-form that admit no Killing vectors. They satisfy the Einstein
field equations with Euclidean signature.

We give the detailed derivation of the solution announced earlier
and present a new solution with an added parameter. We compare our
method of partner symmetries for finding non-invariant solutions
to that of Dunajski and Mason who use `hidden' symmetries for the
same purpose.

\vspace{1mm}



\vspace{1mm}

\section{Introduction}

In this paper we shall present a method for finding non-invariant
solutions of the elliptic complex Monge-Amp\`ere equation,
hereafter to be referred to as $CMA_2$,
\begin{equation}
u_{1\bar 1}u_{2\bar 2} - u_{1\bar 2}u_{\bar 1 2}=1
\label{ma}
\end{equation}
using its symmetries in a non-standard way. This has application
to important problems in physics and mathematics, in particular
the instanton solutions of the Einstein field equations. They are
described by $4$-dimensional K\"ahler metrics
\begin{equation}
d s^2 =  u_{i\bar{k}} \, d z^i d \bar z^k
\label{metr}
\end{equation}
where summation over the two values of both unbarred and barred
indices is understood and subscripts denote partial derivatives.
If the K\"ahler potential satisfies the elliptic complex
Monge-Amp\`ere equation, then the metric satisfies the vacuum
Einstein field equations with Euclidean signature. We shall be
interested in non-invariant solutions of $CMA_2$ which can be used
to construct hyper-K\"ahler metrics without any Killing vectors.
Among them is the $K3$ surface of Kummer which is the most
important gravitational instanton \cite{ahs}.

Recently we suggested that group foliation can serve as a general
method for finding non-invariant solutions of non-linear partial
differential equations \cite{ns,msw}. Historically this method
goes back to the works of Lie \cite{lie} and Vessiot
\cite{vessiot}, see also Ovsiannikov \cite{ovs} for a modern
exposition. For $CMA_2$ group foliation was constructed in
\cite{ns} but due to the complexity of the resolving equations
non-invariant solutions have not yet been obtained in this way.

Therefore in this paper we adopt a different approach which turned
out to be fruitful specifically for $CMA_2$ which we shall call
the method of {\it partner symmetries}, i.e. pairs of symmetries
related by extended Lax equations. Using this method we obtain
non-invariant solutions of the Legendre transform of (\ref{ma}).
The earlier version of it was published in \cite{ms2000} where we
derived linear partial differential equations for a class of
non-invariant solutions of the hyperbolic $CMA_2$. The use of
symmetries here is unlike their standard use in symmetry reduction
\cite{nsmw} which leads to invariant solutions.

We start with the Lax equations \cite{masnew,mw} appropriate to
(\ref{ma}) for a nonlocal complex potential variable $\Phi$.
However, the commutator of the Lax operators does not reproduce
$CMA_2$ itself, but only its differential consequences. In section
\ref{sec-complex} we supplement the Lax pair with another pair of
linear equations so that in the extended linear system $CMA_2$
becomes an {\it algebraic} consequence. The crucial observation
that follows is that this extended system has compatibility
conditions which coincide with the determining equation for
symmetries of $CMA_2$. The complex potential $\Phi$ will therefore
be constructed from the symmetry characteristics \cite{olv} of
$CMA_2$ which are inter-related by the extended Lax equations for
real potentials given in section \ref{sec-real}.

For the real part $\varphi$ of the complex potential we shall use
the translational symmetry of $CMA_2$ and the dilatational
symmetry for its imaginary part $\psi$. In order to finally arrive
at linear equations, having in mind that symmetry characteristics
satisfy linear equations, we choose $\psi$ as a new unknown
instead of $u$ which implies a Legendre transformation. This is
given in section \ref{sec-legendre}.

As a result we arrive at an over-determined set of second order
partial differential equations for the Legendre transform $\psi$
of the unknown $u$. In section \ref{sec-leg2} we discuss its
second and third order differential compatibility conditions.

In section \ref{sec-pot} we integrate the third order differential
compatibility conditions of this system and arrive at six second
order equations satisfied by $\psi$ with coefficients dependent on
$\varphi$ that has no compatibility conditions. In section
\ref{sec-linear} we choose the translational symmetry
characteristic for $\varphi$. This leads to {\it linear} partial
differential equations with constant coefficients for
$v=e^{-\psi}$. We obtain the general solution of these linear
equations.

This solution is related to a particular solution set of the
original equation (\ref{ma}) by a Legendre transformation and
therefore determines a particular set of solutions of the
Euclidean Einstein equations with anti-self-dual Riemann curvature
$2$-form. We shall not need to reconstruct the corresponding
solution set of the original $CMA_2$ equation (\ref{ma}) by means
of the inverse Legendre transformation, but instead make the
Legendre transformation on the metric (\ref{metr}) itself. In
section \ref{sec-riemann} we present the metric. Since the
solution is obviously non-invariant, the corresponding metric has
no Killing vectors. We discuss some properties of the metric. In
particular we show that it saturates Hitchin's bound
\cite{hitchin} between the Euler number and Hirzebruch signature
which means that if the manifold with our metric can be identified
as a compact manifold it will coincide with the $K3$ surface, or a
surface whose universal covering is $K3$. A preliminary version of
this research with less general set of solutions can be found in
\cite{ms} and the announcement of those results is published in
\cite{mns}.

We discuss possible curvature singularities of our metric in
section \ref{sec-singular}. We find that they coincide with those
of the metric. We derive an additional first order partial
differential equation that the potential $v$ must satisfy for the
existence of singularities. We give the general solution for $v$
that gives rise to singularities in the metric and curvature.

Finally, in section \ref{sec-recurs} we establish the relationship
between our approach of partner symmetries to that developed
recently by Dunajski and Mason \cite{dunman,duma} who suggest
invariance with respect to `hidden' symmetries as a method for
obtaining non-invariant solutions  and apply it to Plebanski's
second heavenly equation. For comparison, we construct non-local
recursion operators for symmetries of $CMA_2$ and show that
partnership between local symmetries of this equation is
equivalent to the invariance of solutions of $CMA_2$ with respect
to non-local symmetries of a very special form such that $CMA_2$
itself becomes a consequence of this invariance. The idea that
invariance with respect to non-local `potential' symmetries can
give rise to non-invariant solutions of partial differential
equations has appeared for the first time in the papers of Bluman
and Kumei (see \cite{blum} and references therein).

\section{Complex potential}
\label{sec-complex}
 \setcounter{equation}{0}

In our approach we start with the Lax equations discovered by
Mason and Newman \cite{masnew,mw}
\begin{eqnarray}
\Phi_1 &=& u_{1 \bar 1} \Phi_{\bar 2} - u_{1 \bar 2} \Phi_{\bar 1}
\nonumber
\\ \Phi_2 &=& u_{2 \bar 1} \Phi_{\bar 2} - u_{2 \bar 2} \Phi_{\bar 1}
\label{lax}
\end{eqnarray}
where $\Phi$ is a complex-valued function of its arguments $\{
z^i, \bar z^k \}$ and we skip the spectral parameter as
unnecessary for our purposes. The commutator of the corresponding
Lax pair
\[ [\partial_1+u_{1\bar 2}\partial_{\bar 1}-u_{1\bar 1}\partial_{\bar 2}\;,
\;\partial_2+u_{2\bar 2}\partial_{\bar 1}-u_{2\bar
1}\partial_{\bar 2}] =0\]
does not reproduce $CMA_2$ but only its
differential consequences resulting in the equation
\[ u_{1\bar 1}u_{2\bar 2} - u_{1\bar 2}u_{\bar 1 2}=k\]
where $k$ is an arbitrary real constant with only $3$ inequivalent
choices $k=1$, $k=-1$ and $k=0$. Hence this Lax pair does not
distinguish between elliptic, hyperbolic, or homogeneous $CMA_2$
which is certainly its drawback.

Therefore we supplement the Lax equations (\ref{lax}) with two
more linear equations
\begin{eqnarray}
\Phi_{\bar 1} &=& u_{2 \bar 1} \Phi_1 - u_{1 \bar 1} \Phi_2 \nonumber\\
\Phi_{\bar 2} &=& u_{2 \bar 2} \Phi_1 - u_{1 \bar 2} \Phi_2
\label{laxad}
\end{eqnarray}
such that $CMA_2$ itself emerges as an algebraic compatibility
condition of any three of these equations and also of the complex
conjugate equations for $\bar\Phi$. Alternatively, if we impose
$CMA_2$ independently, then the additional pair of equations
(\ref{laxad}) follows from the original system (\ref{lax}) and
$CMA_2$.

The differential compatibility condition of equations (\ref{lax})
taken in the form $(\Phi_1)_2=(\Phi_2)_1$ and a similar condition
for the complex conjugate system have the form of the determining
equations for symmetry characteristics of $CMA_2$
\begin{equation}
\Box(\Phi)=0, \qquad \Box(\bar\Phi)=0
\label{desym}
\end{equation}
where $\Box$ is the real linear differential operator
\begin{equation}
\Box=u_{2\bar 2}D_1D_{\bar 1}+u_{1\bar 1}D_2D_{\bar 2} - u_{2\bar
1}D_1D_{\bar 2}-u_{1\bar 2}D_2D_{\bar 1} \label{sq}
\end{equation}
and similarly for the system (\ref{laxad}) and its complex
conjugate system. Here $D_i$ are operators of total
differentiation with respect to $z^i$.

The crucial consequence of (\ref{desym}) that we shall exploit in
this paper is that the complex potential $\Phi$ will be
constructed from the symmetry characteristics of $CMA_2$.

  On the other hand, if we consider the second order derivatives
of the K\"ahler potential as unknowns in the system (\ref{lax}),
(\ref{laxad}), then its matrix has the rank $3$. Thus we may solve
for the three derivatives
\begin{eqnarray}
u_{1\bar 1}\!&\!=\!&\!\frac{\Phi_1\Phi_{\bar 1}}{\Phi_2 \Phi_{\bar
2}}\,u_{2\bar 2}+\frac{\Phi_1}{\Phi_{\bar 2}} -\frac{\Phi_{\bar
1}}{\Phi_2} \label{u_1b2}
\\ u_{1\bar 2}\!&\!=\!&\!\frac{\Phi_1}{\Phi_2}\,u_{2\bar 2}
-\frac{\Phi_{\bar 2}}{\Phi_2}\,,\quad u_{2\bar 1}=\frac{\Phi_{\bar
1}}{\Phi_{\bar 2}}\,u_{2\bar 2} +\frac{\Phi_2}{\Phi_{\bar 2}}
\nonumber
\end{eqnarray}
and these expressions satisfy $CMA_2$ (\ref{ma}) identically.
Substituting the expressions (\ref{u_1b2}) into the equations
complex conjugate to (\ref{lax}), (\ref{laxad}) we obtain four
equations with the only one unknown $u_{2\bar 2}$ which give four
different expressions for the same unknown. We need all of these
expressions to coincide which leads to a single algebraic
compatibility condition
\begin{equation}
\Phi_1\bar\Phi_2 - \Phi_2\bar\Phi_1= \bar\Phi_{\bar 1} \Phi_{\bar
2} - \bar\Phi_{\bar 2} \Phi_{\bar 1} \label{or1}
\end{equation}
of the original linear system and its complex conjugate.

Finally, we find that the metric coefficients in (\ref{metr}) can
be expressed  through the complex potential
\begin{equation}
\label{uibj} u_{i\bar k}=\frac{\Phi_i\bar\Phi_{\bar k
}+\bar\Phi_i\Phi_{\bar k} }{\Phi_1\bar\Phi_2-\bar\Phi_1\Phi_2}
\end{equation}
which identically satisfy $CMA_2$ and reality conditions for $u$
on account of (\ref{or1}). We can use (\ref{uibj}) in the
determining equations (\ref{desym}) for symmetries of $CMA_2$ in
order to express them solely through the potentials $\Phi$ and
$\bar\Phi$
\begin{eqnarray}
 & & (\Phi_2\bar\Phi_{\bar 2}+\bar \Phi_2\Phi_{\bar 2})\Phi_{1\bar 1}
+ (\Phi_1\bar\Phi_{\bar 1}+\bar \Phi_1\Phi_{\bar 1})\Phi_{2\bar 2}
- (\Phi_2\bar\Phi_{\bar 1}+\bar \Phi_2\Phi_{\bar 1})\Phi_{1\bar 2}
\nonumber \label{Phiequ}
\\ & &
- (\Phi_1\bar\Phi_{\bar 2}+\bar \Phi_1\Phi_{\bar 2})\Phi_{2\bar 1}
= 0
\end{eqnarray}
and the complex conjugate equation. This is a system of coupled
second-order quasi-linear equations. If we consider the equations
(\ref{lax}), (\ref{laxad}) and their complex conjugates as
B\"acklund transformations, then the system of equations
(\ref{or1}), (\ref{Phiequ}) and the complex conjugate of the
latter equation containing only potentials $\Phi, \bar\Phi$ form
the B\"acklund transform of the complex Monge-Amp\`ere equation
\cite{ms2000}.

\section{Partner symmetries}
\label{sec-real}
 \setcounter{equation}{0}

We shall work with real symmetry characteristics of $CMA_2$ which
are the real and imaginary parts of the potential $\Phi$. So we
set
\[ \Phi = \varphi + i \psi \] whereby (\ref{lax}), (\ref{laxad})
and their complex conjugates become
\begin{eqnarray}
\varphi_1=i(u_{1\bar 1}\psi_{\bar 2}-u_{1\bar 2}\psi_{\bar
1})\nonumber
\\ \varphi_2=i(u_{2\bar 1}\psi_{\bar 2}-u_{2\bar 2}\psi_{\bar 1})
\label{phi}
\end{eqnarray}
and
\begin{eqnarray}
\psi_1=-i(u_{1\bar 1}\varphi_{\bar 2}-u_{1\bar 2}\varphi_{\bar
1})\nonumber
\\ \psi_2=-i(u_{2\bar 1}\varphi_{\bar 2}-u_{2\bar 2}\varphi_{\bar 1})
\label{psi}
\end{eqnarray}
together with their complex conjugate equations. $CMA_2$ is again
an algebraic consequence of any three equations chosen from
(\ref{phi}) and (\ref{psi}) while the system (\ref{psi}) follows
from (\ref{phi}) plus $CMA_2$.

The differential compatibility conditions
$(\varphi_1)_2=(\varphi_2)_1$ and similarly for $\psi$ are
$\Box(\psi)=0$ and $\Box(\varphi)=0$ which again shows that
$\varphi$ and $\psi$ are symmetry characteristics of $CMA_2$. As a
consequence they will satisfy the non-linear first order
compatibility condition
\begin{equation}
\psi_1 \varphi_2 - \varphi_1 \psi_2  = \varphi_{\bar 1} \psi_{\bar
2}  - \psi_{\bar 1} \varphi_{\bar 2} \label{ord1}
\end{equation}
which is (\ref{or1}). We shall call any pair of symmetries
$\varphi$ and $\psi$ related by equations (\ref{phi}) and
(\ref{psi}) {\it partner symmetries} of $CMA_2$.

\section{Legendre transformation and dilatational symmetry}
\label{sec-legendre}
 \setcounter{equation}{0}

We start with the general symmetry generator of $CMA_2$ \cite{bw}
\[\! X\! =\! i(\Omega_1\partial_2-\Omega_2\partial_1
-\Omega_{\bar 1}\partial_{\bar 2}+\Omega_{\bar 2}\partial_{\bar
1}) +C_1 (z^1\partial_1+\bar z^1\partial_{\bar 1}+u\partial_u) +
iC_2 (z^2\partial_2-\bar z^2\partial_{\bar 2}) + H\partial_u \]
where $\Omega(z^i, \bar z^k)$ and $H(z^i, \bar z^k)$ are arbitrary
solutions of the linear system
\[ \Omega_{1\bar 1}=0, \quad \Omega_{2\bar 2}=0, \quad \Omega_{1\bar 2}=0,
\quad \Omega_{2\bar 1}=0 \] and $C_1$ and $C_2$ are real
constants. The corresponding symmetry characteristic \cite{olv}
has the form
\begin{eqnarray}
\widehat\eta & = & i(u_1\Omega_2-u_2\Omega_1+u_{\bar
2}\Omega_{\bar 1} -u_{\bar 1}\Omega_{\bar 2}) \label{sym}\\
&& +C_1 (u-z^1u_1-\bar z^1u_{\bar 1}) -iC_2 (z^2u_2-\bar
z^2u_{\bar 2}) + H \nonumber
\end{eqnarray}
and $\varphi$, $\psi$ can be chosen as special cases of the
expression (\ref{sym}). We note that the choice of $\varphi$ and
$\psi$ as $\Omega$ and $H$ respectively, leads to the metric for
flat space and hence is trivial.

In the following we shall consider in detail the simplest
non-trivial case when $\varphi$ is identified with the symmetry
characteristic independent of $z^1, \bar z^1$ and $ u_2, u_{\bar
2}$
\begin{equation}
\varphi = u_1\omega(z^2)+u_{\bar 1}\bar\omega(\bar z^2) + h(z^2) +
\bar h(\bar z^2) \label{phiexpr}
\end{equation}
which is a general form of such a symmetry following from
(\ref{sym}). We shall find that in order to end up with linear
equations with constant coefficients $\omega$ must be a constant
and without loss of generality we may choose $h$ as a linear
function, so that (\ref{phiexpr}) becomes
\begin{equation}
\varphi = u_1+u_{\bar 1} + \nu(z^2+\bar z^2)
\label{fi}
\end{equation}
where $\nu$ is an arbitrary real constant.

We shall choose for $\psi$ the characteristic of the dilatational
symmetry
\begin{equation}
\psi=u-z^1u_1-\bar z^1u_{\bar 1}
\label{dilat}
\end{equation}
and in order to arrive at linear equations we shall not substitute
({\ref{fi}), (\ref{dilat}) into the systems (\ref{phi}) and
(\ref{psi}) directly. Instead, we regard $\psi$ as a new unknown
replacing the K\"ahler potential $u$. Then we recognize in the
formula (\ref{dilat}) a part of the Legendre transformation
\begin{eqnarray}
 & \psi=u-z^1u_1-\bar z^1u_{\bar 1},\quad
u=\psi-p\psi_p-\bar p\psi_{\bar p}  & \label{leg}
\\ & z^1=-\psi_p,\quad \bar z^1=-\psi_{\bar p},\quad
u_1=p, \quad u_{\bar 1}=\bar p, & \nonumber
\end{eqnarray}
where $z^2$ remains unchanged and $\psi$ and $\varphi$ are now
regarded as functions of $p$, $\bar p$, $z^2$ and $\bar z^2$. This
is therefore a partial Legendre transformation. We note that
\begin{equation}
\psi_{pp}\psi_{\bar p\bar p} - \psi_{p\bar p}^2 \not= 0
\label{exist}
\end{equation}
is the existence condition for the Legendre transformation
(\ref{leg}).

\section{Legendre transform of the basic equations and their
compatibility conditions}
\label{sec-leg2}
\setcounter{equation}{0}

After the Legendre transformation (\ref{leg}) the system of eight
basic equations (\ref{phi}), (\ref{psi}) and their complex
conjugates is linear in the second derivatives of $\psi$ and has
rank five. Hence it can be solved with respect to five second
derivatives $\psi_{pp}, \psi_{p\bar 2}, \psi_{2\bar 2}, \psi_{\bar
p\bar p}, \psi_{\bar p 2}$ in the form
\begin{equation}
\psi_{pp} = A \psi_{p\bar p},\quad \psi_{p\bar 2} = C \psi_{p\bar
p},\quad \psi_{2\bar 2} = B \psi_{p\bar p} \label{psieq}
\end{equation}
together with their complex conjugates. The only remaining second
derivative $\psi_{p\bar p}$ is regarded as parametric. In
equations (\ref{psieq}) the coefficients depend on the first
derivatives of $\varphi$ and $\psi$
\begin{eqnarray}
 A & = & \frac{\varphi_p^2+\psi_p^2+i(\varphi_p\psi_2-\varphi_2\psi_p)}{
\varphi_p\varphi_{\bar p}+\psi_p\psi_{\bar p}}, \nonumber
\\ C & = & \frac{\varphi_p\varphi_{\bar 2}+\psi_p\psi_{\bar 2}
+i(\varphi_{\bar p}\psi_p-\varphi_p\psi_{\bar p})}{
\varphi_p\varphi_{\bar p}+\psi_p\psi_{\bar p}}, \label{coef}
\\ B & = &  \frac{\varphi_2\varphi_{\bar 2}+\psi_2\psi_{\bar 2}
+i(\psi_p\varphi_{\bar 2}-\varphi_p\psi_{\bar
2}+\psi_2\varphi_{\bar p} -\varphi_2\psi_{\bar
p})}{\varphi_p\varphi_{\bar p}+\psi_p\psi_{\bar p}} \nonumber\\
 & = & | A |^2 + | C |^2 -1 \nonumber
\end{eqnarray}
where we have not yet made any choice of $\varphi$.

It is remarkable that two more second derivatives $\psi_{p2}$ and
$\psi_{\bar p\bar 2}$ are cancelled in the equations (\ref{psieq})
which allows us to end up with linear equations.

We note that after the Legendre transformation $CMA_2$ given by
(\ref{ma}) takes the form
\begin{equation}
\psi_{p\bar p}\psi_{2\bar 2}-\psi_{p\bar 2}\psi_{\bar p2} =
\psi_{pp}\psi_{\bar p\bar p} - \psi_{p\bar p}^2 \label{legma}
\end{equation}
which is identically satisfied as a consequence of (\ref{psieq}).
Symmetry characteristic $\varphi$ of $CMA_2$ (\ref{legma})
satisfies the determining equation which is a linearization of
(\ref{legma})
\[ \psi_{\bar p\bar
p}\varphi_{pp}+\psi_{pp}\varphi_{\bar p\bar p} -(2\psi_{p\bar
p}+\psi_{2\bar 2})\varphi_{p\bar p} +\psi_{\bar p2}\varphi_{p\bar
2}+\psi_{p\bar 2}\varphi_{\bar p2} -\psi_{p\bar p}\varphi_{2\bar
2} = 0 , \]
or
\begin{equation}
\bar A\varphi_{pp}+A\varphi_{\bar p\bar p}-(B+2)\varphi_{p\bar p}
+\bar C\varphi_{p\bar 2}+C\varphi_{\bar p2}-\varphi_{2\bar 2}=0
\label{legdesym}
\end{equation}
on account of equations (\ref{psieq}).

The differential compatibility conditions for the system
(\ref{psieq}) have the form
\begin{equation}
(\psi_{pp})_{\bar 2}=(\psi_{p\bar 2})_p,\quad (\psi_{p\bar
2})_2=(\psi_{2\bar 2})_p
\label{compat}
\end{equation}
and two complex conjugate conditions. They involve four third
derivatives which are determined by differentiating (\ref{psieq})
\begin{equation}
\psi_{p p \bar p}=\frac{ A \bar A_p + A_{\bar p} }{ 1 - | A |^2
}\,\psi_{p\bar p} \; , \quad \psi_{p \bar p 2}= \left( \bar C \,
\frac{ A \bar A_p + A_{\bar p} }{ 1 - | A |^2  } +  \bar C_p
   \right) \psi_{p \bar p} \label{psi3rd}
\end{equation}
and their complex conjugates. For example, we differentiate the
first one of the equations in (\ref{psieq}) with respect to $\bar
p$, use the complex conjugate equation and derive the equation for
$\psi_{pp\bar p}$ as follows
\begin{eqnarray*}
 & & \psi_{p p \bar p}=(\psi_{pp})_{\bar p}=(A \psi_{p\bar p})_{\bar p}=
A (\psi_{\bar p\bar p})_p + A_{\bar p}\psi_{p\bar p}
\\ &=& A (\bar A \psi_{p\bar p})_p
+ A_{\bar p}\psi_{p\bar p}= | A |^2 \psi_{pp\bar p} + ( A \bar A_p
+ A_{\bar p})\psi_{p\bar p}
\end{eqnarray*}
and having determined the third derivatives in this way, we find
that the compatibility conditions (\ref{compat}) result in
\begin{eqnarray}
& AC_{\bar p}-CA_{\bar p}-C_p+A_{\bar 2}=0, & \label{comp1} \\
& \bar A A_p+\bar C C_p -A_{\bar p}-C_2=0 \nonumber
\end{eqnarray}
together with their complex conjugates.

\section{Equations for potentials without \newline compatibility conditions}
\label{sec-pot}
 \setcounter{equation}{0}

In equation (\ref{phiexpr}) we suggested the general form of the
symmetry $\varphi$ which has the Legendre transform
\begin{equation}
\varphi = p\,\omega(z^2) + \bar p\,\bar\omega(\bar z^2) + h(z^2) +
\bar h(\bar z^2) \label{legphi}
\end{equation}
but so far we have not used it. Now we want to show that even
keeping a more general form of $\varphi$, our goal of arriving at
linear equations with constant coefficients fixes the final choice
of $\varphi$ as the Legendre transform of (\ref{fi}).

However, in order to make sure that $\varphi$ will be a symmetry
of the Legendre transform (\ref{legma}) of $CMA_2$ and to provide
a democracy between $\psi$ and $\varphi$ we shall impose on
$\varphi$ equations of the same form (\ref{psieq}) as for $\psi$.
We note that our particular choice (\ref{legphi}) of $\varphi$
will satisfy these equations identically and that this also agrees
with the determining equation for symmetries of (\ref{legma})
since then the equation (\ref{legdesym}) becomes the identity $
|A|^2+ | C |^2 -B-1=0$ in (\ref{coef}). As a consequence all our
second order compatibility conditions (\ref{comp1}) are satisfied
and the system (\ref{psieq}) is compatible in the sense that the
equations (\ref{compat}) are identically satisfied and the same
equations for $\varphi$ are satisfied as well.

Thus, for $\psi$ we have the system of the second order equations
(\ref{psieq}) and the third order equations (\ref{psi3rd}). The
coefficients of the latter equations are logarithmic derivatives
of $\varphi_p\varphi_{\bar p} +\psi_p\psi_{\bar p}$\,, so that
these equations take the form
\begin{eqnarray*}
 & &\frac{(\psi_{p\bar p})_p}{\psi_{p\bar p}} =
\frac{A\bar A_p+A_{\bar p}}{1- |A|^2} =
\frac{(\varphi_p\varphi_{\bar p}+\psi_p\psi_{\bar p})_p}{
\varphi_p\varphi_{\bar p}+\psi_p\psi_{\bar p}}
\\ & & \frac{(\psi_{p\bar p})_2}{\psi_{p\bar p}} =
\bar C \, \frac{A\bar A_p+A_{\bar p}}{1- |A|^2}+\bar C_p =
\frac{(\varphi_p\varphi_{\bar p}+\psi_p\psi_{\bar p})_2}{
\varphi_p\varphi_{\bar p}+\psi_p\psi_{\bar p}}
\end{eqnarray*}
and the integrated equations become an additional second order
equation
\begin{equation}
\psi_{p\bar p}= C_1(\varphi_p\varphi_{\bar p}+\psi_p\psi_{\bar p})
\label{eqpsiadd}
\end{equation}
equivalent to the third order equations (\ref{psi3rd}). Here $C_1$
is a real integration constant. In the case $\varphi$ is not
fixed, these equations hold for $\varphi$ as well where we
exchange $\varphi$ to $\psi$ and $C_1$ to $C_2$.

Hence $\psi$  satisfies the system of six equations (\ref{psieq}),
(\ref{eqpsiadd})
\begin{eqnarray}
 & &\psi_{p\bar p}= \varphi_p\varphi_{\bar p}+\psi_p\psi_{\bar p}
\nonumber
\\ & &\psi_{pp} =
\varphi_p^2+\psi_p^2+i(\varphi_p\psi_2-\varphi_2\psi_p)
\label{psisys}
\\ & &\psi_{p\bar 2} = \varphi_p\varphi_{\bar 2}+\psi_p\psi_{\bar
2} +i(\varphi_{\bar p}\psi_p-\varphi_p\psi_{\bar p}) \nonumber
\\ & &\psi_{2\bar 2}
= \varphi_2\varphi_{\bar 2}+\psi_2\psi_{\bar 2}
+i(\psi_p\varphi_{\bar 2}-\varphi_p\psi_{\bar
2}+\psi_2\varphi_{\bar p} -\varphi_2\psi_{\bar p}) \nonumber
\end{eqnarray}
together with two complex conjugate equations and similarly for
$\varphi$. Here we have used (\ref{eqpsiadd}) in the system of
equations (\ref{psieq}) and the expressions (\ref{coef}) for the
coefficients $A$, $B$ and $C$ and made the change of notation
$C_1\psi\mapsto\psi$, $C_1\varphi\mapsto\varphi$. The differential
compatibility conditions of these systems are identically
satisfied without generating any second, or third order
conditions. Thus the Legendre transform of the determining
equation (\ref{desym}) for symmetries of $CMA_2$ is also
identically satisfied for $\psi$ and $\varphi$ together with the
first order compatibility condition (\ref{ord1}). Therefore, any
$\varphi$ and $\psi$ which satisfy the systems (\ref{psisys}) and
the one obtained from (\ref{psisys}) by the exchange of $\psi$ and
$\varphi$ form a pair of partner symmetry characteristics related
to each other by the Legendre transform of (\ref{phi}) and
(\ref{psi}).

\section{Linear equations and their general solution}
\label{sec-linear}
 \setcounter{equation}{0}

The linearization is achieved by the logarithmic substitution
\begin{equation}
\psi=-\log{v} \label{vv}
\end{equation}
 so that the first equation in (\ref{psisys})
\[-\frac{v_{p\bar p}}{v}+\frac{v_pv_{\bar p}}{v^2} =
\varphi_p\varphi_{\bar p}+\frac{v_pv_{\bar p}}{v^2}\]
becomes the linear equation
\begin{equation}
v_{p\bar p} = - \varphi_p\varphi_{\bar p}\,v
\label{lineq}
\end{equation}
because according to our choice (\ref{phiexpr}) $\varphi$ is
independent of $z^1, \bar z^1$ and $ u_2, u_{\bar 2}$. Indeed if
we had admitted such dependence, then after the Legendre
transformation we would have $z^1=-\psi_p$, $u_2=\psi_2$ so that
the derivatives of $\varphi$ in (\ref{lineq}) and (\ref{psisys})
would result in non-linear dependence on $v$. Furthermore, the
second derivatives $v_{p2}, v_{\bar p \bar 2}$ would enter
destroying the whole structure.

The linear equation (\ref{lineq}) will have a constant coefficient
if we further impose the condition $\varphi_p=const$. That is, we
choose the symmetry $\varphi$ to be linear in $p,\bar p$ which
results in the Legendre transform of the formula (\ref{fi})
\begin{equation}
\varphi = p+\bar p +\nu (z^2+\bar z^2) \label{fileg}
\end{equation}
where $\nu$ is an arbitrary real constant and our choice of
$h(z^2)=\nu z^2$ in (\ref{phiexpr}) has been made without loss of
generality. Thus, from now on we fix the choice of the potential
$\varphi$ as the characteristic of translational symmetry
(\ref{fi}) with its Legendre transform (\ref{fileg}).

Then the equations (\ref{psisys}) for $v=e^{-\psi}$ become linear
with constant coefficients
\begin{eqnarray}
 & & v_{p\bar p} + v = 0 \nonumber
\\ & & v_{pp} + v - i(v_2-\nu v_p) = 0
\label{linv}
\\ & & v_{p\bar 2} + \nu v - i(v_p-v_{\bar p}) = 0\nonumber
\\ & & v_{2\bar 2} + \nu^2 v - i\bigl[v_2-v_{\bar 2}
+\nu(v_p-v_{\bar p})\bigr] = 0 \nonumber
\end{eqnarray}
plus two complex conjugate equations.

The general solution of the linear system (\ref{linv}) in the case
of the discrete spectrum has the form
\begin{eqnarray}
\!\!\! &\!\!\! & v = \sum_{j=-\infty}^{\infty} \exp
\left\{2\,{\rm Im}\! \left(\frac{}{} \left[\alpha_j^2(s_j^2+1)+1
\right]z^2 \right) \right\} \left\{ \frac{}{} \right.
\label{solut}
\\ \!\!\! &\!\!\! &\exp\!\left[\frac{}{} 2s_j {\rm Re} [\alpha_j (p+\nu z^2) ] \right]\!
{\rm Re}\!\left\{\frac{}{}\!F_j \exp\!\left[ 2 i\,\Bigl[{\rm
Im}\bigl(\alpha_j(p+\nu z^2) \bigr) - 2s_j {\rm Re}(\alpha_j^2
z^2) \Bigr] \right]  \right\} + \nonumber
\\ \!\!\! &\!\!\! & \left.
\exp\!\left[\frac{}{}\!\! -2 s_j {\rm Re}[\alpha_j (p+\nu z^2) ]
\right]\! {\rm Re}\!\left\{\frac{}{}\!\! G_j \exp\!\left[2
i\,\Bigl[{\rm Im}\bigl(\alpha_j(p+ \nu z^2) \bigr) + 2s_j {\rm
Re}(\alpha_j^2 z^2) \Bigr] \right]\! \right\} \!\!
 \right\} \nonumber
\end{eqnarray}
where $\alpha_j,F_j, G_j$ are arbitrary complex constants and
$s_j=\sqrt{1-1/|\alpha_j|^2}$. For a particular example one can
take a finite sum instead of infinite series in this formula. On
the other hand we can also consider the continuous spectrum where
$\alpha_j$ should be changed to $\alpha$, $s_j$ to
$s=\sqrt{1-1/|\alpha|^2}$, $F_j, G_j$ to $F(\alpha,\bar\alpha),
G(\alpha,\bar\alpha)$ respectively and the sum in the formula
(\ref{solut}) should be changed to a double integral with respect
to $\alpha,\bar\alpha$. In this way we end up with an integral
representation of the solution for $v$. Since the solution
(\ref{solut}) explicitly depends on four real independent
variables, it is a non-invariant solution of the Legendre
transform of $CMA_2$ given by (\ref{legma}).

For our particular purpose of constructing the metric without any
Killing vectors we have no need to reconstruct the corresponding
non-invariant solution of the original field equation (\ref{ma}),
which is $CMA_2$ itself, by the inverse Legendre transformation.
Instead we make the Legendre transformation (\ref{leg}) in the
metric (\ref{metr}) itself.

\section{The metric}
\label{sec-riemann}
\setcounter{equation}{0}

The solution (\ref{solut}) of the linear equations (\ref{linv})
will be used in the construction of the metric which is a Legendre
transform of the metric (\ref{metr}) . For this purpose it will be
convenient to introduce a new notation for the numerators and the
denominator of the coefficients $A$ and $C$ in (\ref{coef})
putting $A =a/c, \; C = \bar b / c$
\begin{eqnarray}
 a & = &  v^2 + v_p^2 - iv(v_2-\nu v_p), \nonumber\\
 b & = & v_{\bar p} v_2 + \nu v^2 - iv(v_p-v_{\bar p}),
 \label{abdnew}\\
 c & = & v^2 + | v_p |^2 \nonumber
\end{eqnarray}
so that the metric is given by
\begin{eqnarray}
ds^2 &=& \frac{1}{ v^2 (c^2 - |a|^2 ) } \left[ \frac{}{} a \, ( c
\, d p + b \, d z^2)^2 +\bar a \, ( c \, d \bar p+  \bar b \,
d\bar z^2)^2\right.
\label{newmetr}
\\ & & \left.\mbox{}+\frac{1}{c}\,( c^2 + | a |^2) | \, c \, d p
+ b \, d z^2 \, |^2 \right] + \frac{1}{ v^2 c} (c^2 -|a|^2) | d
z^2 |^2 \nonumber
\end{eqnarray}
with the real potential $v$ determined by (\ref{solut}). The
solution (\ref{solut}) depends on four independent variables, so
that it is a non-invariant solution. The Legendre-transformed
hyper-K\"ahler metric (\ref{newmetr}) therefore has no Killing
vectors. This general result is violated in only one case, if the
sum in (\ref{solut}) is restricted to only one term. We note that
the metric coefficients depend only on logarithmic derivatives of
$v$ and therefore in this special case the dependence on the
argument of the first exponential factor in (\ref{solut}) vanishes
and the metric depends on only three coordinates. This is a
symmetry reduction.

Any solution for $v$ of the form (\ref{solut}) with a minimum of
two terms in the sum when substituted into (\ref{abdnew}) gives us
an explicit form of the metric (\ref{newmetr}) without any Killing
vectors.

It will be useful to express the metric (\ref{newmetr}) in the
Euclidean Newman-Penrose formalism \cite{g1}, \cite{an}. The
metric is given by
\begin{equation}
d s^2 =  l \otimes \bar l + \bar l \otimes l +
  m \otimes \bar m + \bar m \otimes m  \label{np}
\end{equation}
where
\begin{eqnarray} l &=& \frac{1}{v\left[  \frac{}{} c \, ( c^2 -|a|^2 )
\right]^{1/2}}   \left[ \, c \, ( c \, d p + b \, d z^2 ) + \bar a
\, ( c \, d \bar p + \bar b \, d \bar z^2)  \frac{}{} \right] , \nonumber \\
m&=& \frac{\left( c^2 -| a |^2 \right)^{1/2}}{v \, c^{1/2}} d z^2
\label{npframe}
\end{eqnarray}
and the co-frame will be labelled as $\omega^a = \{ l, \bar l, m,
\bar m \}$.

It can be verified directly that in the Newman-Penrose frame with
the metric coefficients given by (\ref{abdnew}) and the potential
$v$ satisfying the linear system (\ref{linv}), the Riemann
curvature $2$-form is anti-self-dual
\begin{equation}
\Omega^a_{\;b} = - ^* \Omega^a_{\; b}, \hspace{1cm} \Omega^a_{\;
b} = \frac{1}{2} R^a_{\;bcd} \; \omega^c \wedge \omega^d
\label{anti}
\end{equation}
where $^*$ is the Hodge star operator. Ricci-flatness follows by
virtue of the first Bianchi identity. The metric (\ref{newmetr})
has no Killing vectors since the potential $v$ in the solution
(\ref{solut}) depends on all four coordinates. Its first Chern
class vanishes since $R_{i k} \, \omega^i \wedge \omega^k =0$.
There are three real closed $2$-forms that follow from the metric
(\ref{newmetr})
\begin{eqnarray}
\theta_0 &=& l \wedge \bar l - m \wedge \bar m \nonumber \\
\theta_+   &=& \frac{1}{2} \left(  l \wedge \bar m + \bar l \wedge
m \right) \label{hyper} \\
\theta_-   &=& \frac{1}{2 i} \left( l \wedge \bar m - \bar l
\wedge m  \right) \nonumber
\end{eqnarray}
which shows that it is hyper-K\"ahler because the $(1,1)$ tensors
that define the structure functions of three almost complex
structures are obtained by raising an index of the $2$-forms
(\ref{hyper}) with the metric \cite{an}.

We note that due to the fact that $v$ is given by exponentials and
the metric coefficients are homogenous of degree zero in $v$ and
its derivatives, the metric coefficients tend to constant values
asymptotically. The same is true for the Newman-Penrose tetrad
scalars of the Riemann tensor.

We shall not discuss whether, or not our solution describes a
compact $4$-manifold. All our analysis has been local and given a
metric in a local coordinate chart as in (\ref{newmetr}),
compactness is always an open question. The property of
compactness depends on the range of co-ordinates that we may
assign to the local coordinates $\{p, \bar p, z^2, \bar z^2 \}$.
We have not done that, but the presence of exponentials in the
metric coefficients suggests that the metric could well be made
compact by choosing a suitable domain of coordinates. Assuming
compactness, by virtue of the anti-self-dual curvature property,
our solution saturates Hitchin's bound $|\tau|\le (2/3)\chi$
\cite{hitchin} between the Euler characteristic $\chi$ and the
Hirzebruch signature $\tau$. They are defined as integrals of the
following $4$-forms over a compact manifold
\[ \chi =\frac{1}{2^4\pi^2}\int \Omega^a_{\;b}\wedge
{^*\Omega}^b_{\; a}, \quad \tau = \frac{1}{24\pi^2}\int
\Omega^a_{\;b}\wedge \Omega^b_{\; a} \] and hence the quantity
\[\chi + \frac{3}{2}\,\tau = \frac{1}{2^4\pi^2}\,\int
\Omega^a_{\;b} \wedge \left({^*\Omega}^b_{\; a} + \Omega^b_{\;
a}\right) = 0
\]
vanishes due to the anti-self-duality (\ref{anti}) of the Riemann
curvature $2$-form. Thus we have the saturation of the Hitchin
bound. Only $K3$ and surfaces whose universal covering is $K3$
have this property according to Hitchin's theorem \cite{hitchin}.

The metric (\ref{newmetr}) contains an infinite number of
arbitrary parameters, even arbitrary functions of two variables in
the double integral. Furthermore, it is an open question as to
whether, or not for suitably chosen values of the arbitrary
constants and domain of coordinates this metric will be that of a
compact manifold. This is an important issue, because of its
relevance to the metric on $K3$. We know from index theorems
\cite{as} that the number of parameters in the metric for $K3$ is
finite. Further work will possibly enable us to identify the
essential parameters in this solution that actually characterize
$K3$.

\section{Analysis of possible curvature singularities}
\label{sec-singular}
\setcounter{equation}{0}
In order to discuss singularities in the curvature scalars we will
need the tetrad scalars of the Riemann tensor in the
Newman-Penrose frame (\ref{npframe}). The Newman-Penrose tetrad
scalars of the Riemann tensor are too lengthy to be presented
here. However, their denominators are quite simple and the only
place where curvature scalars blow up are given by
\begin{equation}
|a|^2 - c^2 =0 \label{singnew}
\end{equation}
which are also singularities of the metric. Using (\ref{abdnew})
we arrive at the following first order partial differential
equation
\begin{equation}
 v \left[ \frac{}{} (v_p - v_{\bar p} )^2  +| v_2 - \nu v_p |^2 \frac{}{}
 \right] + 2\,{\rm Im}\!\left[\frac{}{}(v_2 -\nu v_{p})(v^2 + v_{\bar p}^2)
 \frac{}{}\right] = 0
 \label{sing}
\end{equation}
for possible curvature singularities. If for some $v$ the equation
(\ref{sing}) is satisfied together with the system (\ref{linv}),
then the denominator of the metric is identically zero and hence
such solution for $v$ can not be used for constructing the metric.
From the definitions (\ref{abdnew}) and the singularity condition
(\ref{singnew}) we note the following relations
\begin{eqnarray}
 & & c_p = i(b-\nu c),\quad a_{\bar p} = -i(b-\nu c),\quad
a_p = i\,\frac{a}{c}\,(b-\nu c) \label{rel}
\\ & & \frac{b}{c} - i\left(\frac{v_{\bar p}}{v}\,\frac{a}{c} - \frac{v_p}{v}
\right) -\nu = 0,\qquad \frac{\bar b}{c}-\nu =
\frac{c}{a}\left(\frac{b}{c} - \nu \right) \label{relat}
\end{eqnarray}
which are used to check that $ A_p = A_{\bar p} = 0,\; C_p =
C_{\bar p} = 0$.

Next we express $v_{p 2}$ from the last equation (\ref{rel}) and
check the relation
\begin{equation}
i\left(v_{p2}-v_{\bar p 2}\,\frac{a}{c}\right) +
v_2\left(\frac{b}{c}-\nu \right) = 0
\label{help}
\end{equation}
by differentiating the first one of equations (\ref{relat}) with
respect to $z^2$ and use the relation (\ref{help}). We find
\begin{equation}
\left(\frac{b}{c}\right)_2 - i\,\frac{v_{\bar
p}}{v}\left(\frac{a}{c}\right)_2 = 0 \label{res}
\end{equation}
and differentiating this equation with respect to $p$, using the
independence of $A =a/c $ and $\bar C =b/c $ on $p$ we prove their
independence of $z^2, \bar z^2$ as well by checking that $A_2 =
A_{\bar 2} = 0,\; C_2 = C_{\bar 2} = 0$ and hence $A$ and $C$ are
constants
\begin{eqnarray}
& \displaystyle A =\frac{a}{c}=\lambda=const,\quad \bar C
=\frac{b}{c}=\mu=const &
\label{const}
\\[2pt] &\displaystyle \bar\lambda = \frac{1}{\lambda}\,,\quad
\bar\mu = \frac{{\mu+(\lambda-1)\nu}}{\lambda} & \nonumber
\end{eqnarray}
which serve as the definition of $\lambda$ and $\mu$. From the
definitions of $a$ and $c$ it follows that
\[v (c - a) + iv_p (b - \nu c) - i (v_2-\nu v_p) c = 0 \]
which due to (\ref{const}) becomes
\begin{equation}
v_2 = \mu v_p + i(\lambda -1)v \label{v_2}
\end{equation}
and the relation (\ref{help}) takes the form
\begin{equation}
v_p = \lambda v_{\bar p} + i(\mu-\nu)v .
\label{v_p}
\end{equation}
Compatibility conditions $(v_p)_2=(v_2)_p$ and $(v_p)_{\bar
2}=(v_{\bar 2})_p$ of (\ref{v_p}), (\ref{v_2}) and the complex
conjugate to (\ref{v_2}) are identically satisfied.

Consecutive integration of these three first order equations gives
the solution
\begin{eqnarray}
v & = & \exp{\left\{ i(\lambda -1)\left(z^2+\frac{1}{\lambda}\,\bar
z^2\right) +i(\mu-\nu)\xi\right\}} H(\eta) \label{presol} \\
\xi &=&p+ \mu z^2 , \qquad \qquad \eta=\lambda\xi+\bar\xi
\nonumber
\end{eqnarray}
Due to the first of the second-order equations (\ref{linv}) the
function $H(\eta)$ satisfies the following ordinary differential
equation
\begin{equation}
\lambda H^{\prime\prime} + i(\mu-\nu) H^\prime + H = 0 \label{ode}
\end{equation}
with the general solution
\begin{eqnarray}
H & = & C_1\exp{\left\{\frac{i}{2\lambda}\left[-(\mu-\nu) +
\sqrt{(\mu-\nu)^2+4\lambda} \,\right] \eta\right\}} \nonumber
\\ & + & \bar C_1 \exp{\left\{ -\frac{i}{2\lambda}
\left[\mu-\nu + \sqrt{(\mu-\nu)^2+4\lambda}\,\right] \eta\right\}}
\label{H}
\end{eqnarray}
where $C_1, \bar C_1$ are arbitrary constants and the second
constant was chosen as $\bar C_1$ to satisfy the reality condition
\[e^{i(\mu-\nu)\xi}H=e^{-i\,\frac{(\mu-\nu)}{\lambda}\,\bar\xi}\bar H\]
for the solution $v$ defined by (\ref{presol}). In the notation
\[\lambda=-\frac{\alpha}{\bar \alpha}\,,\quad \mu=\nu - 2i\alpha s,\quad
s=\sqrt{1-\frac{1}{|\alpha|^2}},\quad C_1 = \bar F \]
the singular solution (\ref{presol}) with $H$ defined by (\ref{H}) finally
becomes
\begin{eqnarray}
& & v = {\rm Re}\!\left\{ F \exp \left\{ \frac{}{}
\alpha(s+1)p+\bar\alpha(s-1)\bar p
 \right. \right. \label{finsol}\\
& & \left. \left. - i \! \left[\alpha^2(s+1)^2+1 + i
\nu\alpha(s+1) \right]z^2 + i\! \left[\bar\alpha^2(s-1)^2+1 -
i\nu\bar\alpha(s-1)\right]\bar z^2 \! \right\} \! \right\}
\nonumber
\end{eqnarray}
where $F$ is an arbitrary complex constant. The expression
(\ref{finsol}) for $v$ is the general solution of the system of
linear second order equations (\ref{linv}) supplemented by the
first order singularity condition (\ref{sing}). It is obvious that
the singular solution (\ref{finsol}) is a very special case of the
general solution (\ref{solut}) which is obtained if we restrict
ourselves to only one term in the sum (\ref{solut}) and keep only
one complex constant $F$ putting the other one $G$ equal to zero.
This solution is automatically avoided since the minimum number of
terms in the sum (\ref{solut}) is two in order not to have
symmetry reduction.

Thus, we have proved that any solution (\ref{solut}) of the linear
system (\ref{linv}) which does not coincide with (\ref{finsol})
can be used to construct the explicit expression for metric
(\ref{newmetr}).

\section{Recursion operators and invariance with respect to non-local symmetries}
\label{sec-recurs}
\setcounter{equation}{0}

In this section we establish the relationship between our use of
partner symmetries for obtaining non-invariant solutions of
$CMA_2$ to the method of Dunajski and Mason \cite{dunman,duma}.
Dunajski and Mason have suggested a general approach for finding
non-invariant solutions as invariant solutions with respect to
`hidden' symmetries and applied it to the second heavenly equation
of Plebanski.

We introduce a pair of linear differential operators
\begin{equation}
L_1 = i(u_{1\bar 2}D_{\bar 1}-u_{1\bar 1}D_{\bar 2}),\quad L_2 =
i(u_{2\bar 2}D_{\bar 1}-u_{2\bar 1}D_{\bar 2}) \label{Lop}
\end{equation}
plus their complex conjugates. These operators commute as a
consequence of $CMA_2$, alternatively $[ L_1, L_2] =0 $ implies
differential consequences of $CMA_2$. In terms of these operators
the relations (\ref{phi}) and (\ref{psi}) between potentials take
the form
\begin{eqnarray}
\varphi_1 = -L_1 \psi , & \quad & \varphi_2 = -L_2 \psi ,
\label{fiLpsi}
\\ \psi_1 = L_1 \varphi, & \quad & \psi_2 = L_2 \varphi
 \nonumber
\end{eqnarray}
The determining equation $\Box(\varphi)=0$ for symmetries of
$CMA_2$ with the operator $\Box$ given by (\ref{sq}) can be
expressed in terms of the operators (\ref{Lop})
\begin{equation}
L_2 D_1 \varphi = L_1 D_2 \varphi\quad \iff \quad D_1 L_2 \varphi
= D_2 L_1 \varphi \label{deLop}
\end{equation}
Equation (\ref{deLop}) is a conservation law which implies the
existence of a potential $\psi$ for the symmetry $\varphi$ such
that
\begin{equation}
D_1 \psi = L_1 \varphi, \quad D_2 \psi = L_2 \varphi
\label{psipot}
\end{equation}
and we note that $\psi$ satisfies the same equation $\Box(\psi)=0$
taken in the form (\ref{deLop})
\[D_1 L_2 \psi = D_2 L_1 \psi \quad \Longrightarrow \quad
L_2 D_1 \psi = L_1 D_2 \psi\ \quad \Longrightarrow \quad
L_2L_1\varphi = L_1L_2\varphi \] where we have used equations
(\ref{psipot}). The last equation is satisfied identically since
the operators $L_1$ and $L_2$ commute on the solution manifold of
$CMA_2$.

Hence we find that if $\varphi$ is a symmetry of $CMA_2$ then
$\psi$ defined by
\begin{equation}
\psi = D_1^{-1} L_1 \varphi = D_2^{-1} L_2 \varphi \label{defom}
\end{equation}
is also a symmetry. Therefore the integro-differential operators
\begin{equation}
R_1 = D_1^{-1} L_1, \quad R_2 = D_2^{-1} L_2 \label{recurs}
\end{equation}
are recursion operators for symmetries  of $CMA_2$ defined on the
subspace of symmetries $\varphi$ satisfying the relations
$R_1\varphi=R_2\varphi$.

The relations (\ref{fiLpsi}) between the potentials become
\begin{equation}
\varphi = -R_1\psi,\quad \varphi = -R_2\psi, \qquad \psi =
R_1\varphi,\quad \psi = R_2\varphi \label{relpot}
\end{equation}
so that comparison of (\ref{defom}) with the two latter formulas
in (\ref{relpot}) shows that our second symmetry $\psi$ which is a
partner for $\varphi$ coincides with the potential for $\varphi$.
Hence the recursion operators $R_i$ are defined on the subspace of
partner symmetries of $CMA_2$.

The operators $R_i$ are recursion operators for the whole space of
symmetries of $CMA_2$ if we define the inverse integral operators
$D_i^{-1}$ by specifying the limits of integration. The relations
$D_i^{-1}D_i=1$ for $i=1,2$ will be satisfied if we define
$D_i^{-1}=\int_{a}^{z^i} dz^i$ and impose the boundary conditions
$f(a)=0$ for all $f$ in the domain of definition of $D_i^{-1}$. In
particular, if $a=\infty$ we need to restrict ourselves to the
space of functions decreasing to zero when $z^i$ tend to infinity.
Then we have
\[D_i^{-1} D_i f = \int\limits_{a}^{z^i}D_i(f(z^i)) dz^i = f(z^i) \]
for all such $f$ and hence $D_i^{-1}D_i=1$.

Recursion operators $R_i$ should commute with the operator
(\ref{sq}) of the determining equation for symmetries of $CMA_2$
on the space of its solutions on account of $CMA_2$. Calculating
these commutators we obtain
\[ [R_i,\Box] = [L_i,D_i^{-1}]\, \Box \]
where $$\Box=i(D_2L_1 - D_1L_2)$$ coincides with the operator
(\ref{sq}). Hence on the space of solutions of the equation
$\Box(\varphi) = 0$ we have the required property
$[R_i,\Box]\varphi = 0$.

If we choose for $\varphi$ and $\psi$ {\it any} local symmetries
of $CMA_2$, then the following expressions are characteristics of
non-local symmetries
\begin{equation}
\widehat\eta_1 = R_1\psi + \varphi,\quad \widehat\eta_2 = R_2\psi
+ \varphi,\quad \widehat\eta_3 = R_1\varphi - \psi, \quad
\widehat\eta_4 = R_2\varphi - \psi
\label{nonloc}
\end{equation}
so that the relations (\ref{relpot}) of partnership between local
symmetries take the form of equations determining invariant
solutions of $CMA_2$ with respect to the non-local symmetries
(\ref{nonloc})
\begin{equation}
 \widehat\eta_i = 0,\qquad i = 1,2,3,4
\label{inv}
\end{equation}
of a very particular type (\ref{nonloc}). Taking into account
$CMA_2$, only two relations out of the system (\ref{phi}),
(\ref{psi}) are independent and therefore the same is true for the
invariance conditions (\ref{inv}). Alternatively, $CMA_2$ itself
is an algebraic consequence of the invariance conditions
(\ref{inv}). We note that the definition of $D_i^{-1}$ so that
$R_i$ become recursion operators on the space of all symmetries of
$CMA_2$ is not needed after we impose the condition (\ref{inv})
because it ensures that $R_i$ now act only on the subspace of
partner symmetries.

 Thus, the starting point of our use of local partner symmetries of
$CMA_2$ is equivalent to searching for its solutions invariant
with respect to two non-local symmetries of a special form
(\ref{nonloc}) which is in the spirit of Dunajski and Mason's
treatment of Plebanski's second heavenly equation. Both of these
approaches exploit the idea that an appropriate non-standard use
of symmetries of a non-linear partial differential equation may
result in a construction of its non-invariant solutions. However,
the use of symmetries is very different in these two methods. The
resulting linear equations are therefore also completely
different.

For the sake of completeness we note that `hidden' symmetries,
recursion operators and infinite hierarchies of the
self-dual-gravity equations were also studied in \cite{pop,lecht}.

\section{Conclusion}

We have shown that a class of non-invariant solutions of $CMA_2$
can be obtained by solving a set of linear partial differential
equations with constant coefficients for a single real potential.
This has enabled us to obtain explicit expressions for the
Legendre transform of hyper-K\"ahler metrics with anti-self-dual
Riemann curvature $2$-form that admit no continuous symmetries.
From the anti-self-duality property it follows that their first
Chern class vanishes and the Einstein field equations with
Euclidean signature are satisfied.

We started with the extension of Mason-Newman Lax equations for
the complex potential of $CMA_2$ and found that the equation
determining the symmetry characteristics of $CMA_2$ appeared as
their integrability condition which lead us to the concept of
partner symmetries. The use of partner symmetries proved to be the
basic tool for finding non-invariant solutions of $CMA_2$ by
solving linear equations.

We established the relation between our concept of partner
symmetries and invariance of solutions of $CMA_2$ with respect to
such non-local symmetries that $CMA_2$ itself becomes a
consequence of this invariance. This seems to be in the spirit of
the idea of using `hidden' symmetries for finding non-invariant
solutions of the second heavenly equation suggested in the recent
work of Dunajski and Mason. However, our use of symmetries and
linear equations determining non-invariant solutions of $CMA_2$ is
completely different.

In this paper we used the translational and dilatational
symmetries of $CMA_2$ as partner symmetries. We plan to use other
pairs of partner symmetries in a future publication in order to
obtain further classes of non-invariant solutions of $CMA_2$ and
the corresponding metrics without continuous symmetries.

\section{Acknowledgement}

We thank Professor L. Nirenberg for encouragement. When he learned
that we were working on finding explicit non-invariant solutions
of $CMA_2$, he said that he would be surprised but then kindly
added that he had been surprised before.

This work was supported in part by T\"UBA, Turkish Academy of
Sciences.

\end{document}